\documentclass[seceq]{iis}
\usepackage{graphics}
\usepackage{color}
\usepackage{amsmath}
\usepackage{setspace}
\usepackage{caption}

\title{Quantum information and statistical mechanics: an introduction to frontier}
\runtitle{quantum information and statistical mechanics}

\author{%
\name{Keisuke \surname{FUJII}}$^{1,2,3}$
\CAE{keisukejayorz@gmail.com}, 
}
\inst{
$^{1}$
The Hakubi Center for Advanced Research, Kyoto University, Yoshida-Ushinomiya-cho, Sakyo-ku, Kyoto 606-8302, Japan 
\\
$^{2}$
Graduate School of Informatics, 
Yoshida Honmachi, Sakyo-ku, Kyoto 606-8501, Japan 
\\
$^{3}$
Graduate School of Engineering Science, Osaka University,
Toyonaka, Osaka 560-8531, Japan \\

}
\runauthor{Keisuke Fujii}

\support{Grant-in-Aid for Scientific Research on Innovative Areas No. 20104003.}

\recdate{20YY}{0}{DD}%% set by the editors
\accdate{20YY}{0}{DD}%% set by the editors

\abst{This is a short review on an interdisciplinary field of quantum information science
and statistical mechanics.  We first give a pedagogical introduction to the stabilizer formalism,
which is an efficient way to describe an important class of quantum states,
the so-called stabilizer states,
and quantum operations on them. 
Furthermore, graph states,
which are a class of stabilizer states associated with graphs, and their applications 
for measurement-based quantum computation are also mentioned.
Based on the stabilizer formalism, 
we review two interdisciplinary topics. One is the relation between
quantum error correction codes and spin glass models,
which allows us to analyze the performances of quantum error correction codes
by using the knowledge about phases in statistical models.
The other is the relation between the stabilizer formalism and 
partition functions of classical spin models,
which provides new quantum and classical algorithms to evaluate
partition functions of classical spin models.}

\kword{\kw{quantum information science}, \kw{stabilizer formalism}, \kw{statistical mechanics}}

\begin{document}
\maketitle

\section{Introduction}
\label{Introduction}
Quantum information science is a rapidly growing field of physics,
which explores comprehensive understanding of extremely complex quantum systems.
An ultimate goal is building quantum computer or quantum simulator,
which outperform conventional high performance (classical) computing.
To achieve this goal,
there still exist a lot of problems to be overcome.
One of the main obstacles is decoherence, i.e.,
loss of coherence, due to an undesirable interaction between environments.
Without a fault-tolerant design, any promising computing scheme is nothing but
pie in the sky. Actually it has been shown that NP-complete problems 
(even PSPACE-complete problems),
which are though to be intractable in classical digital computer, 
can be solved in polynomial time 
by using analog computer that can perform 
$x+y$, $x-y$, $xy$, and $\lfloor x \rfloor$ for any two real numbers $x$ and $y$
\cite{Aaronson,Schonhage}.
However, such an analog computer has not been realized
so far, since unlimited-precision real numbers cannot be physically realizable.
Fortunately, in quantum computation, we have a fault-tolerant theory
\cite{Shor95,DiVincenzo-Shor96,Kitaev97a,Preskill98,Knill98,Knill98b,Aharanov-Ben-Or99,Aharanov-Ben-Or08},
which guarantees an arbitrary accuracy of quantum computation
as long as noise levels of elementary quantum gates are sufficiently
smaller than a constant value, the so-called threshold value.
Fault-tolerant quantum computation utilizes
quantum error correction \cite{Shor95} to overcome decoherence,
where we utilize many physical particles and classical processing to infer
the error location.
Another important problem for quantum computation
is to identify the class of problems that can be solved by using quantum computer. Of course, we already have several quantum algorithms,
Shor's factorization \cite{Shor94}, Grover's search \cite{Grover}, etc., which outperform existing classical algorithms.
However, these instances are not enough 
to understand the whole class of problems 
that quantum computer can solve, i.e., BQP problems,
since they are, so far, not shown to be BQP-complete.
Finding a good BQP-complete problem will lead us
to a deeper understanding.

Interestingly, in both fault-tolerant quantum computing
and finding new quantum algorithms,
the idea of statistical mechanics has been 
applied, recently. 
One is a correspondence between 
quantum error correction and spin glass theory~\cite{Dennis},
where posterior probabilities in Bayesian inference problems
for quantum error correction are mapped onto
partition functions of spin glass theory.
Thus the knowledge about phases in spin glass theory
tells us the performance of quantum error correction codes~\cite{Wang,Ohzeki,Roth,FujiiTokunaga,OhzekiFujii,Al-Shimary,Loss1,Loss2,OhzekiLoss}.
Another is a relationship between
overlaps (inner products) of quantum states
and partition functions of statistical mechanical models~\cite{Bravyi,Nest1,Completeness}.
Since the overlaps (inner products) can be estimated efficiently in quantum computer,
this mapping gives us a new quantum algorithm
to calculate the partition functions.
Are these correspondences between quantum information
and statistical mechanics accidental or inevitable?
At least there is one common idea behind them,
stabilizer states or stabilizer formalism~\cite{Gottesman},
which might be an important clew to bridge the two fields
and find new things.

The stabilizer states are 
a class of quantum states,
which takes quite important roles in quantum information processing.
The stabilizer formalism provides an efficient tool to provide 
a class of quantum operations on stabilizer states.
In this review, we give a pedagogical introduction
to the stabilizer formalism. Then we review 
two interesting interdisciplinary topics
between quantum information science and 
statistical mechanics: the correspondence between quantum error correction 
and spin glass theory and the relation between the stabilizer formalism and partition functions of statistical models.

The rest of the paper is organized as follows:
In Sec. 2, we introduce quantum-bit and 
elementary gates (unitary matrices).
In Sec. 3, we introduce the stabilizer formalism.
In Sec. 4, we explain how to describe a class of unitary operations 
and measurements on the stabilizer states in the stabilizer formalism.
In Sec. 5, we demonstrate utility of the stabilizer formalism
on the graph states~\cite{Graph}, which are also an important class of quantum states
associated with mathematical graphs. Specifically,
we explain how Pauli basis measurements transform the graph states.
Furthermore, we also introduce a model of quantum computation, measurement-based quantum computation.
In Sec. 6, we review the relation between quantum error correction codes
and spin glass models. In the last part of Sec. 6,
we also mention a route toward building a fault-tolerant quantum computer.
In Sec. 7, we review the relation between 
the stabilizer formalism and partition functions of classical spin models.
As an exercise, we also show a duality relation
between two distinct spin models by using the stabilizer formalism.
Section 8 is devoted to a conclusion.
 
\section{Quantum bit and quantum gates}
In classical information science,
the minimum unit of information is 
described by a binary digit or {\it bit},
which takes a value either 0 or 1.
Its quantum counterpart
is a quantum bit, so-called {\it qubit}.
Qubit is defined as a superposition of 
two orthogonal quantum states $|0\rangle$ and 
$|1\rangle$,
$|\psi\rangle = \alpha |0\rangle + \beta |1\rangle$,
where $\alpha$ and $\beta$ are
arbitrary complex values satisfying
$|\alpha|^2 + |\beta|^2=1$.
An $n$-qubit state is described by 
\begin{eqnarray*}
| \Psi \rangle = \sum _{i_1 , i_2 ,... , i_n} C_{i_1 i_2 ... i_n} | i_1 i_2 ... i_n\rangle,
\label{eq01}
\end{eqnarray*}
where $| i_1 i_2 ... i_n\rangle \equiv |i_1 \rangle \otimes |i_2 \rangle \otimes ... \otimes |i_n\rangle$.
Since time evolutions are given by unitary operators
in quantum physics,
gate operations for a single qubit can be given by
$2 \times 2$ unitary matrices.
The most fundamental gates 
 are identity and Pauli operators:
\begin{eqnarray*}
I
=
\left(
\begin{array}{cc}
1 & 0
\\
0 & 1
\end{array}
\right),
X
=
\left(
\begin{array}{cc}
0 & 1
\\
1 & 0
\end{array}
\right),
Y
=
\left(
\begin{array}{cc}
0 & -i
\\
i & 0
\end{array}
\right),
Z
=
\left(
\begin{array}{cc}
1 & 0
\\
0 & -1
\end{array}
\right).
\end{eqnarray*}
The set of tensor products of 
the Pauli matrices on the $n$-qubit system 
$ \{ \pm 1, \pm i \} \times \{ I , X, Y, Z \}^{\otimes n}$
forms the $n$-qubit Pauli group $\mathcal{P}_{n}$.
%
%These $X$, $Y$, $Z$ and $-iI$
%generate a group,
%so-called {\it Pauli group}.
A Pauli operator $A \in \{ X, Y, Z\}$ acting on
the $i$th qubit is denoted by $A_i \equiv I _1 \otimes  ... \otimes I_{i-1} 
\otimes A_i \otimes   I_{i+1}  \otimes ... \otimes  I_n$.

In stead of the computational basis $\{ |0\rangle ,|1\rangle \}$, 
we may also choose different orthogonal 
bases:
\begin{eqnarray*}
\{ |+\rangle \equiv (|0\rangle + |1\rangle)/\sqrt{2},
|-\rangle \equiv (|0\rangle - |1\rangle)/\sqrt{2}
\},
\textrm{ or }
\{ |+i \rangle \equiv (|0\rangle +i |1\rangle)/\sqrt{2},
|-i\rangle \equiv (|0\rangle -i |1\rangle)/\sqrt{2}
\}.
\end{eqnarray*}
The Hadamard and phase gates
are defined as
\begin{eqnarray*}
H=
\frac{1}{\sqrt{2}}
\left(
\begin{array}{cc}
1 & 1
\\
1 & -1
\end{array}
\right)
\textrm{ and }
S=
\left(
\begin{array}{cc}
1 & 0
\\
0 & i
\end{array}
\right).
\end{eqnarray*}
These gates 
transform the computational bases,
$\{ |0\rangle , |1\rangle\}
\leftrightarrow \{ |+\rangle, |-\rangle \}$
and
$\{ |+ \rangle , |-\rangle\} \leftrightarrow \{ |+i \rangle, |-i\rangle\}$,
respectively.
These $X$, $Y$, $Z$, $H$ and $S$ gates
are normalizers of the Pauli group
and generate a group $\mathcal{C}_{1}$of single-qubit Clifford gates.
(That is, a Clifford gate, say $A \in \mathcal{C}_{1}$, transforms 
the single-qubit Pauli group onto itself under the conjugation $A [\cdot ] A^{\dag}$.
For example, $HXH=Z$, $SXS^{\dag} =Y$, and so on.)
Since the single-qubit Clifford group
is a discrete group,
one cannot generate arbitrary 
single-qubit unitary operations.
Fortunately, it has been known that
the existence of a non-Clifford gate in addition to the Clifford gates 
is enough for an efficient construction of an arbitrary single-qubit gate with
arbitrarily high accuracy
according to the Kitaev-Solovay theorem~\cite{KitaevSolovay}.
A non-Clifford gate can be, for example,
$\pi /8 $ gate $U_{\pi /8} = e^{ - i \pi Z /8}$. 

Single-qubit gates are not enough for constructing 
an arbitrary $n$-qubit unitary matrix, and hence
at least one two-qubit gate operation is inevitable.
There are two famous  two-qubit gates,
controlled-NOT (CNOT) and controlled-$Z$ (C$Z$) gates,
which are given, respectively, by
\begin{eqnarray*}
\Lambda (X)_{c,t} = |0\rangle \langle 0| _{c} \otimes 
I_{t} + |1\rangle \langle 1|_{c} \otimes X_{t},
\;\;\;\;
\Lambda (Z)_{c,t} = |0\rangle \langle 0| _{c} \otimes 
I_{t} + |1\rangle \langle 1|_{c} \otimes Z_{t}.
\end{eqnarray*}
Here a controlled-$A$ gate is denoted by $\Lambda (A)_{c,t}$,
where the gate $A$ acts on the state labeled by $t$ 
conditioned on the state labeled by $c$.
Thus these qubits $c$ and $t$ are called control and target qubits, respectively.
These two two-qubit gates are both Clifford gates,
that is, $\Lambda (A)$ ($A=X,Z$) transforms $\mathcal{P}_2$ 
onto itself under the conjugation $\Lambda (A) [ \cdots ] \Lambda (A)^{\dag}$.
For example,
$\Lambda (X) _{c,t} X_c \otimes I_t \Lambda (X) _{c,t}^{\dag}  = X_c \otimes X_t$,
$\Lambda (X) _{c,t} I_c \otimes Z_t  \Lambda (X) _{c,t}^{\dag}= Z_c \otimes Z_t$,
$\Lambda (Z) _{c,t} X_c \otimes I_t\Lambda (Z) _{c,t}^{\dag}  = X_c \otimes Z_t$,
and so on.
It is known that
an arbitrary unitary operation of $n$ qubits
can be constructed from
these elementary gates
$\{ H,S,U_{\pi/8}, \Lambda (X) \}$ 
or $\{ H,S,U_{\pi/8},\Lambda (Z) \}$. 
%\cite{Reck94,DiVincenzo95,Bernstein97}. 
Such a set of elementary unitary gates
is called a {\it universal set},
which is an instruction set of quantum computer.

\section{Stabilizer states}
In general, a description of quantum states is difficult
since it requires exponentially many parameters in the number of qubits
as shown in Eq. (\ref{eq01}).
To understand such a complex quantum system, 
efficient tools to describe important 
classes of complex quantum systems are essential.
The matrix-product-states (MPS)~\cite{Fannes91}, projected-entangled-pair-states (PEPS)~\cite{Verstraete06,Verstraete08,Cirac09},
and multiscale-entanglement-renormalization-ansatz (MERA)~\cite{Vidal} are
such examples. The stabilizer states are 
another important class of quantum states,
which take important roles in quantum information processing.

Now we introduce the definition of stabilizer states and 
the stabilizer formalism~\cite{Gottesman}.
We define a stabilizer group $\mathcal{S}$ of an $n$-qubit system
as an Abelian subgroup of the $n$-qubit Pauli group that does not includes $-I$ 
as its element.
In other words, all elements in the stabilizer group
are commutable with each other and hermitian. 
For example, a set 
\begin{eqnarray*}
\mathcal{S}_{\rm Bell}=\{ II, XX, ZZ, -YY\}
\end{eqnarray*} 
is a two-qubit stabilizer group,
where $A\otimes B$ is denoted by $AB$ for simplicity.

The stabilizer group $\mathcal{S}$ can be defined in terms of the maximum independent set of the stabilizer group, which we call stabilizer generators.
Here independence is defined such that each element in the generator set cannot be expressed as 
a product of other elements in the generator set.
The stabilizer group generated by a stabilizer generators $\{S_i \}$ 
is denoted by $\langle \{ S_i \} \rangle$.
For example, the stabilizer group 
$\mathcal{S}_{\rm Bell}=\{ II, XX, ZZ, -YY\}$ can be written simply by $\langle XX, ZZ \rangle$.
Any stabilizer group of $n$ qubits can be defined
if at most $n$ stabilizer generators are given. 

The stabilizer state is defined as a simultaneous eigenstate
of all stabilizer elements with the eigenvalue +1.
It is sufficient that the state is an eigenstate of 
all stabilizer generators:
\begin{eqnarray*}
S_i | \psi \rangle = |\psi \rangle  \textrm{ for all stabilizer generators } S_i.
\end{eqnarray*}
The dimension $2^d$ of the space spanned by the stabilizer states
is calculated to be $2^d=2^{n-k}$ with $n$ and $k$ being the number of qubits and stabilizer generators,
respectively.
This can be understood that the $2^{n}$-dimensional space
is divided into two orthogonal subspaces for each stabilizer generator.

For example, the stabilizer state defined by the stabilizer group
$\mathcal{S}_{\rm Bell}=\langle XX, ZZ \rangle$ is $(|00\rangle + |11\rangle )/\sqrt{2}$,
which is a maximally entangled state of two qubits~\cite{EPR}.
The maximally entangled state is a useful resource for teleporting
an unknown quantum state between
two separate sites, that is, quantum teleportation~\cite{teleportation}.

Another representative example of the stabilizer states is an $n$-qubit cat state,
\begin{eqnarray*}
| {\rm cat } \rangle = \frac{1}{\sqrt{n}} ( | 00 ... 0 \rangle + |11 ... 1\rangle ),
\end{eqnarray*}
whose stabilizer group is given by 
\begin{eqnarray*}
\left\langle Z_1 Z_2, \; ... , \; Z_{n-1}Z_{n}, \;\prod _{i=1}^{n} X_i \right\rangle,
\end{eqnarray*}
where $A_i$ indicates a Pauli operator $A \in \{ X,Y,Z\}$ on the $i$th qubit.
The cat state is a representative example of macroscopically entangled states.
If one particle is determined whether it is $|0\rangle$ or $|1\rangle$,
the superposition is completely destroyed.

\section{Quantum operations on stabilizer states}
Let us consider the action of a Clifford gate $U$ on the stabilizer state $|\psi \rangle$
defined by a stabilizer group $\mathcal{S}=\langle \{ S_i \} \rangle$:
\begin{eqnarray*}
U | \psi \rangle = US_i |\psi \rangle  = U S_i U^{\dag}U  | \psi \rangle  = S'_i  U| \psi \rangle,
\end{eqnarray*}
where $S'_i \equiv U S_i U^{\dag} $.
This indicates that a state $U | \psi \rangle $ 
is an eigenstate of the operator $S'_i$ with an eigenvalue +1 for all $S'_i$.
Furthermore,
since the Clifford group of unitary gates
are normalizer of the Pauli group (and hence a Pauli product is transformed to another Pauli product
under its conjugation),
the group $\mathcal{S}'= \langle \{S'_i \} \rangle$ is also a stabilizer group.
That is, $U | \psi \rangle $ is the stabilizer state defined by the stabilizer group $\mathcal{S}'$.
In this way, the action of $U$ on the stabilizer state can be regarded by
a map between the stabilizer groups. For example,
a stabilizer group $\langle X_1I_2, I_1Z_2 \rangle$ (the corresponding state is $|+\rangle_1 |0\rangle_2$)
is transformed to $\langle X_1X_2, Z_1Z_2 \rangle$ (the corresponding state is $(|00\rangle  + |11\rangle)/\sqrt{2}$)
by $\Lambda (X)_{1,2}$.

The projective measurement of a hermitian Pauli product as an observable
can also be described by a map between stabilizer groups as follows.
Suppose $A$ is an $n$-qubit hermitian Pauli product and $P_{\pm} = (I \pm A)/2$ are the corresponding projection operators.
Due to the projective measurement, a state $| \psi \rangle$ is 
projected to $P_{+} | \psi \rangle / \sqrt{||  P_+| \psi \rangle ||} $ or 
$P_{-} | \psi \rangle /\sqrt{ || P_-| \psi \rangle ||} $ with probabilities $ || P_{\pm}| \psi \rangle ||$.
For simplicity, we consider the case with the measurement outcome $+$.
There are three possibilities:
\begin{itemize}
\item[(i)] $A$ (or $-A$) is an element of the stabilizer group.
\item[(ii)] While $A$ (or $-A$) does not belong to the stabilizer group,
$A$ is commutable with all elements of the stabilizer group.
\item[(iii)] At least one stabilizer operator does not commute with $A$.
\end{itemize}
In the case (i), the state does not change.
In the case (ii), the stabilizer group for the post-measurement state
is given by $\langle A, \{ S_i \} \rangle$. That is,
$A$ is added to the generator set.
In the case (iii), we can always choose 
a generator set $\langle S_1, S_2 ,... S_k \rangle$  ($k < n$) of the original stabilizer group
such that $A$ commutes with all generators $S_2, ..., S_k$ except for $S_1$.
Then the stabilizer group of the post-measurement state
is given by $\langle A, S_2, ... , S_k \rangle$.
That is, in addition to adding $A$, $S_1$ is removed from the generator set.

%%%%%%%%%%%%%%%%%%
\section{Graph states}
In this section, we exercise to use the stabilizer formalism.
As an example of the stabilizer states,
we use the most important subclass of stabilizer states, graph states~\cite{Graph}.
A graph state is defined associated with a graph $G(V,E)$,
where $V$ and $E$ are the sets of the vertices and edges, respectively.
A qubit is located on each vertex of the graph.
The stabilizer generator of the graph state $|\psi _{G}\rangle$ associated with a graph $G(V,E)$ is 
given by
\begin{eqnarray*}
K_i =  X_i \prod _{ j \in N_i } Z_j \;\; \textrm {for all } i \in V ,
\end{eqnarray*}
where the neighbor $N_i:= \{ j | (i,j) \in E \}$ of vertex $i$ indicates the set of vertices that are connected 
to vertex $i$ on the graph.

Each stabilizer generator $K_i$  
is transformed to $X_i =U K_i U$ by
$U = \prod _{(i,j) \in E} \Lambda (Z)_{i,j}$,
where we aggressively used the fact that $\Lambda (Z) _{i,j} X_i I_j \Lambda (Z)_{i,j}
= X_i Z_j$.
This fact indicates that $U$ transforms 
the stabilizer group from $\langle \{ K_i \}_{ i \in V} \rangle$ to $\langle \{ X_i  \}_{i \in V}\rangle$.
The stabilizer state defined by the latter is $|+ \rangle ^{ \otimes |V|}$ where $|V|$ is the number of the vertices.
This implies that we can obtain the graph state $| \psi _G \rangle $
from the product state $|+\rangle ^{\otimes |V|}$ as $|\psi _G \rangle = U |+ \rangle ^{\otimes |V|}$.
Specifically, when the graphs are regular such as one-dimensional, square, hexagonal, and cubic  lattices,
the corresponding graph states are also referred to as cluster states~\cite{Cluster}.
It has been known that any stabilizer state is equivalent to a certain graph state
up to local Clifford operations~\cite{LCeq,Graph}. However, the graph associated with a stabilizer state
is not uniquely defined, since there exist local Clifford operations that change the associated graph.
This property is called local complementarity of the graph states~\cite{LCeq,Graph}.

\begin{figure}[t]
\centering
\includegraphics[width=120mm,keepaspectratio,clip]{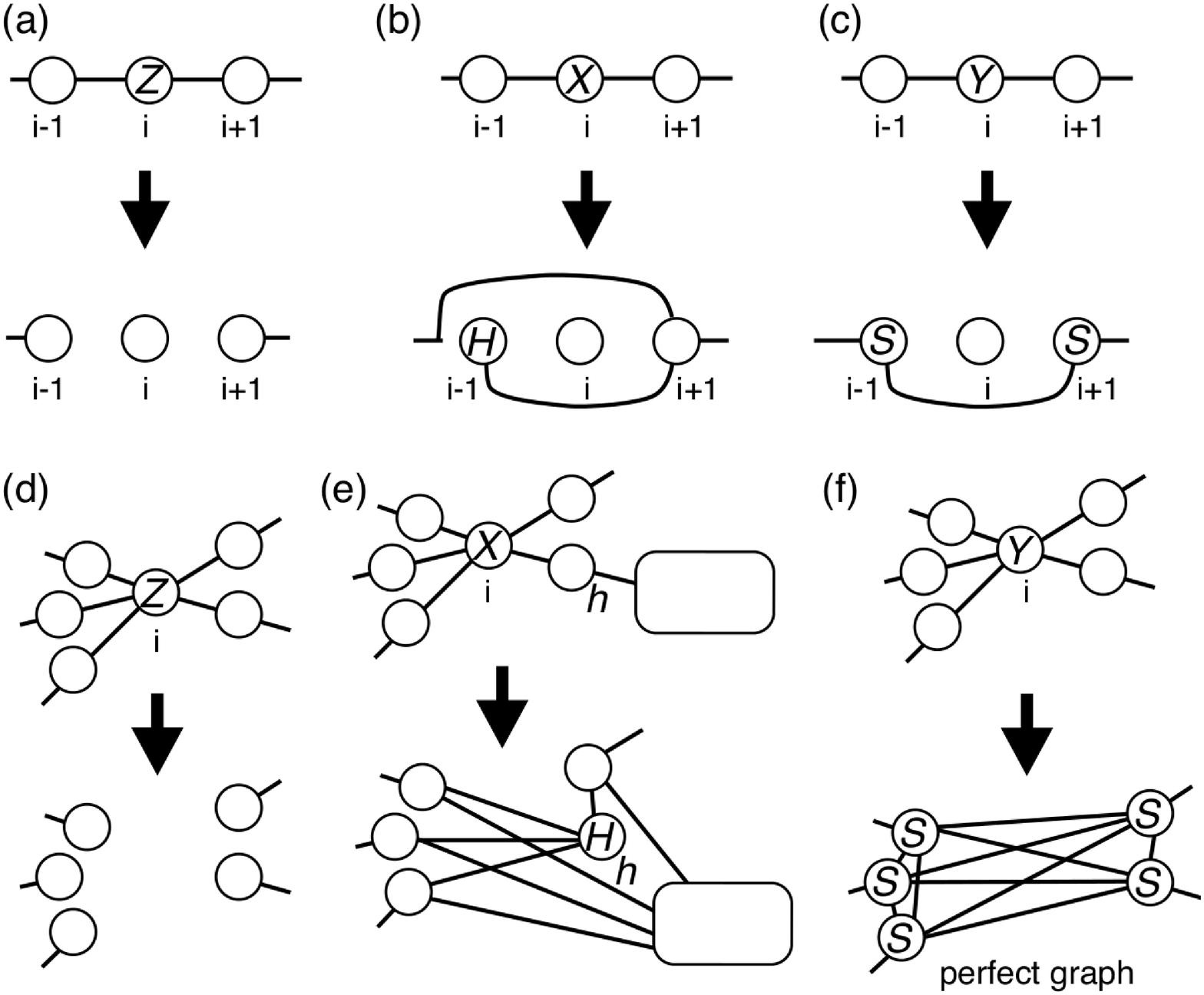}
\caption{The transformations of graph states by the Pauli basis measurements.
(a) The $Z$-basis measurement on a one-dimensional graph state. (b) The $X$-basis measurement on 
a one-dimensional graph state. $H$ indicates the Hadamard gate. (c) The $Y$-basis measurement 
on a one-dimensional graph state. $S$s indicates the phase gates. 
(d) The $Z$-basis measurement
on a general graph state. The edges incident to the $i$th (measured) qubit
is removed. (e) The $X$-basis measurement on a general graph state. A qubit $h$
is chosen from the set $V_i$ of qubits connected to the $i$th (measured) qubit. 
All edges connected to the $i$th qubit are removed, and
the edges between
$V_i \setminus h$ and $h$ and between $V_i \setminus h$ and vertices which are connected with $h$
(shown by a blank box) are added. 
In addition, the $H$ gate is performed on the $h$th qubit.
For arbitrary choices of $h \in V_i$, the associated graph states,
which are equivalent up to local Clifford gates with each other, are defined.
(f) The $Y$-basis measurement on a general graph state. 
All edges connected to the $i$th qubit are removed. 
Instead, a complete graph of the vertices $V_i$ 
adjacent to the $i$th (measured) qubits
is added. In addition, the $S$ gates are performed on the qubits belonging to $V_i$.  }
\label{fig1}
\end{figure}

Next we will see how Pauli basis measurements transform the graph states.
Let us consider a one-dimensional graph state as shown in Figs. \ref{fig1} (a)-(c).
At first, we consider the $Z$ basis measurement (projective measurement of the observable $Z$) 
on the $i$th qubit. The stabilizer group for the post-measurement state
is given by
\begin{eqnarray*}
\langle ..., K_{i -1}, Z_{i}, K_{i+1} ,...\rangle,
\end{eqnarray*}
since $K_i$ does not commute with $Z_i$.
After the projection, the $i$th qubit is $|0\rangle$
and hence decoupled with the other qubits.
By rewriting the stabilizer generators,
we obtain three decoupled stabilizer groups
\begin{eqnarray*}
\langle ..., Z_{i-2}X_{i-1}  \rangle  , \langle Z_i \rangle, \langle   X_{i+1} Z_{i+2} ,...\rangle,
\end{eqnarray*}
which means the graph is separated by the $Z$ basis measurement as shown in Fig.~\ref{fig1} (a).

Similarly, we consider the $X$-basis measurement.
Since $X_i$ does not commute with $K_{i-1}$ and $K_{i+1}$
but commutes with $K_{i-1}K_{i+1}=Z_{i-2} X_{i-1}X_{i+1} Z_{i+2}$,
the stabilizer group for the post-measurement state
is given by
\begin{eqnarray*}
\langle ...,Z_{i-2} X_{i-1}X_{i+1} Z_{i+2}, Z_{i-1}Z_{i+1}  ,...\rangle, \langle X_i \rangle.
\end{eqnarray*}
By performing the Hadamard gate $H$ on the $(i-1)$th qubit,
we obtain a new stabilizer group 
\begin{eqnarray*}
\langle ...,Z_{i-2} Z_{i-1}X_{i+1} Z_{i+2}, X_{i-1}Z_{i+1}  ,...\rangle, \langle X_i \rangle,
\end{eqnarray*}
which indicates that the graph is transformed
as shown in Fig.~\ref{fig1} (b) up to the Hadamard gate.
Instead of the $(i-1)$th qubit,
we can obtain a similar result by performing
the Hadamard gate on the $(i+1)$th qubit.

The final example is the $Y$-basis measurement.
Since $Y_i$ does not commute with $K_{i-1}$, $K_i$ and $K_{i+1}$
but commutes with $K_{i-1}K_{i}= Z_{i-2} Y_{i-1} Y_i Z_{i+1} $,
and $K_{i}K_{i+1}= Z_{i-1} Y_{i} Y_{i+1} Z_{i+2} $,
the stabilizer group for the post-measurement state
is given by
\begin{eqnarray*}
\langle ...,Z_{i-2} Y_{i-1} Z_{i+1}, Z_{i-1} Y_{i+1} Z_{i+2}  ,...\rangle, \langle Y_i \rangle.
\end{eqnarray*}
By performing the phase gates $S$ on the $(i-1)$th and $(i+1)$th
qubits, we obtain a new stabilizer group
\begin{eqnarray*}
\langle ...,Z_{i-2} X_{i-1} Z_{i+1}, Z_{i-1} X_{i+1} Z_{i+2}  ,...\rangle, \langle Y_i \rangle.
\end{eqnarray*}
This indicates that the graph is directly connected up to the phase gates as shown in Fig.~\ref{fig1} (c).

The actions of the Pauli basis measurements for general graph structures
can be also calculated in a similar manner as shown in Fig.~\ref{fig1} (d)-(f),
where no loop is included for clarity, but their extensions to 
arbitrary graphs are straightforward.
\begin{figure}[hbt]
\centering
\includegraphics[width=140mm,keepaspectratio,clip]{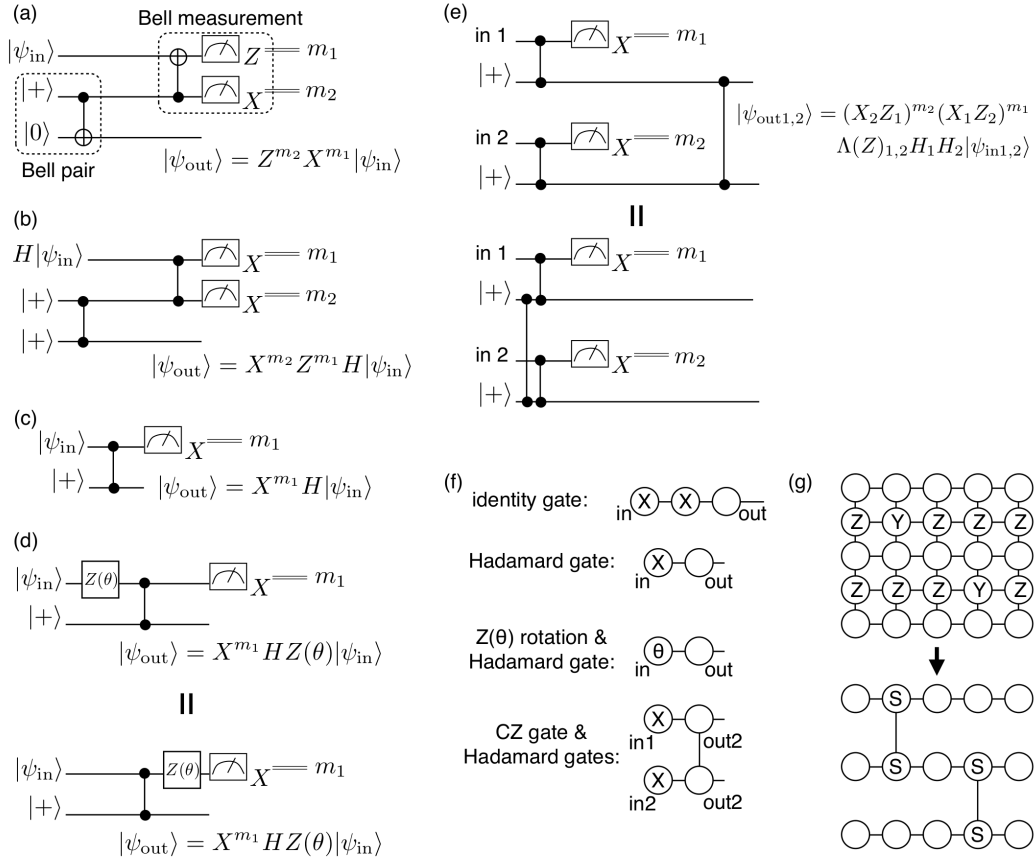}
\caption{(a) A circuit for quantum teleportation \cite{teleportation}. The two-qubit gates are the CNOT gates, where the control and target qubits are denoted by black and white circles, respectively. (b) An equivalent circuit of (a), which consists of the C$Z$ gates.
The Hadamard gates come from the equivalence $\Lambda (X) _{c,t} = 
H_{t} \Lambda (Z)_{c,t} H_{t}$. (c) A minimum unit of teleportation-based gate, which transforms the input state to the output with the Hadamard operation. (d) 
The circuit equivalence for the $HZ(\theta)$ operation.
The measurement in the $X$ basis after the $Z(\theta)$ rotation
teleports the output state with the $HZ(\theta)$ operation 
up to a Pauli byproduct.
(e) The C$Z$ gate for the output states (upper circuit) can be 
moved into the generation of a graph state (lower circuit),
where the commutability of C$Z$ gates is utilized. 
(f) Graph diagrams of MBQC.
From top to bottom, they describe
the identity gate (b), the Hadamard gate (c), the $Z(\theta)$ rotation 
followed by the Hadamard gate (d),
and the teleportation-based C$Z$ gate followed by
the Hadamard gates (e). 
(g) A graph state required for quantum computation
is generated from the 2D cluster state on a square lattice
by using Pauli basis measurements.}
\label{fig4}
\end{figure}

The graph states take important roles in quantum information processing~\cite{Graph}.
First of all, graph states defined on a certain class of graphs, such as square and hexagonal lattices, can be used for 
resources for measurement-based quantum computation (MBQC)~\cite{MBQC1,MBQC2,Universal1,Universal2}.
This can be understand as follows.
The circuit in Fig.~\ref{fig4} (a) is well-known 
quantum teleportation \cite{teleportation}, 
where a Bell pair $(|00\rangle + |11\rangle)/\sqrt{2}$
is prepared as a resource, and the unknown input state is 
teleported into the output up to a Pauli byproduct depending on the 
measurement outcomes.
This circuit is equivalent to that in Fig.~\ref{fig4} (b),
which reads $X$-basis measurements on a graph state 
transfer the input state to the output state.
Here we should recall that a graph state is generated by C$Z$ gates from $|+\rangle ^{\otimes n}$.
The measurement-based identity gate can be decomposed into
two Hadamard gates, one of which is shown in Fig.~\ref{fig4} (c).
By measuring in the $X$ basis after $Z(\theta)\equiv e^{ -i \theta Z/2}$,
a unitary transformation $HZ(\theta)$ can be performed
in a teleportation-based way as shown in circuit equivalences in Fig.~\ref{fig4} (d).
The Pauli byproduct made due to the probabilistic nature of measurements
has to be always placed at the top of the output state to handle the nondeterminism.
However, $Z(\theta)$ does not commute with the Pauli byproduct if it contains the Pauli $X$ operator,
i.e., $Z(\theta)X=XZ(-\theta)$.
To settle this, 
the measurement angle $\theta$ is adaptively changed to $- \theta$ beforehand
according to the previous measurement outcomes,
which is called feedforward in MBQC.

By using commutability of C$Z$ gates as shown in Fig.~\ref{fig4} (e),
the C$Z$ gate performed for the output of teleportation
can be moved into the offline graph state preparation.
This trick is the so-called quantum gate teleportation \cite{GottesmanChuang}.
In this way, an arbitrary single qubit rotation and the C$Z$ gate,
which are sufficient for universal quantum computation,
can be implemented by adaptive measurements
on a specific type of graph state
Starting from, for example, a two-dimensional (2D) cluster state on a square lattice,
we can generate an arbitrary graph state required for universal quantum computation
by using Pauli basis measurements mentioned before.
Roughly speaking, the universality of measurement-based quantum computation
means that by performing measurements in appropriately chosen bases $\{ | \alpha _i \rangle\}$,
the output of a quantum computation $U | 0\rangle ^{ \otimes n}$ of $n$ qubits can be simulated as
\begin{eqnarray}
U | 0\rangle ^{ \otimes n} =  2^{(|V|-n)/2} \left( \bigotimes _{i \in V \verb+\+ \textrm{output $n$ qubits}}\langle \alpha _i | \right) | \psi _G \rangle .
\label{MBQC}
\end{eqnarray}
Such a state, which allows universal quantum computation
in a measurement-based way, is called universal resource.
Recently, MBQC on more general many-body quantum states
has been proposed \cite{Gross1,Gross2},
which utilizes MPS and PEPS as resources.
MBQC on ground or thermal states of valence-bond-solid systems
have been proposed \cite{Brennen,Cai,Bartlett,Wei,Miyake,LiRaussendorf,FujiiMori}, which might be useful to relax experimental difficulties
in preparations of universal resources for quantum computation.

\section{Quantum error correction codes and spin glass models}
In the following two sections, we will review the interdisciplinary topics 
between quantum information science and statistical mechanics. In this section,
we describe the correspondence between quantum error correction codes and spin glass models.

Quantum error correction~\cite{Shor95} is one of the most successful schemes to handle errors
on quantum states originated from an undesirable interaction with environments.
Quantum error correction codes employ multiple physical qubits to encode logical information
into a subspace, so-called {\it code space}.
The stabilizer formalism is vital to describe such a complex quantum system.
Below we will describe one of the most important quantum error correction codes,
the surface code~\cite{Kitaev,Dennis}, which has a close relation to a spin glass model, the random-bond Ising model (RBIM).

\begin{figure}[t]
\centering
\includegraphics[width=130mm,keepaspectratio,clip]{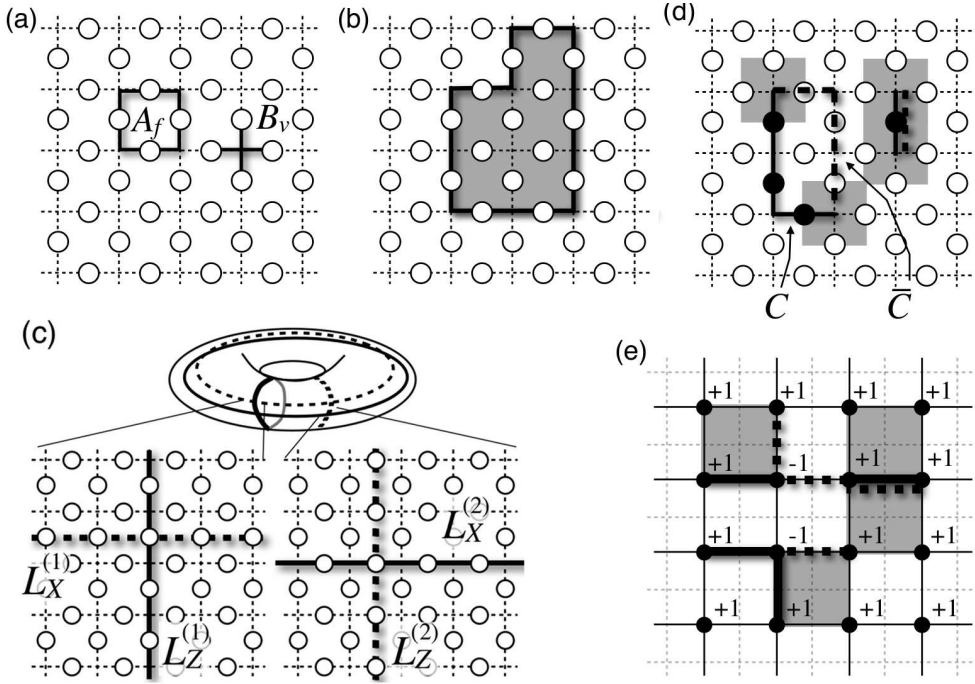}
\caption{
(a) The vertex and face stabilizer operators. 
(b) A trivial cycle. The product
of the Pauli $Z$ operators on a trivial cycle is equivalent to 
the product of face stabilizer generators 
inside the cycle, and hence it belongs to the stabilizer group.
(c) Two pairs of the logical operators,
{\tiny $L_{X}^{(1)}$} (left dotted line), {\tiny $L_{Z}^{(1)}$} (left solid line), {\tiny$L_X^{(2)}$} (right dotted line), 
and {\tiny $L_Z^{(2)}$} (right solid line). 
(d) The black circles and solid lines denote a chain $C$ of $Z$ errors. The vertices with $b_v = -1$ are denoted by 
gray squares. The estimated chain $\bar C$ of errors for a recovery is shown by dotted lines. If $C+\bar C$ is a trivial cycle, the error correction succeeds.
Otherwise if $C+\bar C$ is a nontrivial cycle,
it changes the code space nontrivially. (e) RBIM on the dual lattice. The error chain $C$ corresponds
to the locations of antiferromagnetic interactions. The recovery chain $\bar C$ 
corresponds to the excited domain wall. The vertices with $b_v=-1$ correspond
to Ising vortexes. At zero temperature,
a spin configuration is chosen so that the length of excited domain wall is minimum,
i.e., MWPM of the Ising vortexes (or frustrations).}
\label{fig2}
\end{figure}

The surface code on the square lattice consists of 
qubits located on edges $i \in E$ of the square lattice~\cite{Kitaev,Dennis}
(note that the location of qubits is different from that of the graph states).
Specifically, we consider a periodic boundary condition,
that is, the surface code on a torus.
The face and vertex stabilizer operators of the surface code
are defined for each face $f \in F$ and vertex $v \in V$ as
\begin{eqnarray*}
A_f=\prod _{ i \in E_f } Z_{i}, 
\textrm{ and   }\;\;
B_v=\prod _{j \in E_v} X_j,
\end{eqnarray*}
respectively
 [see Fig.~\ref{fig2} (a)].
%Here, $Z_i$ and $X_j$ denote
%Pauli operators on the $i$th and $j$th qubits ($i,j \in E$),
%
Here $E_{f}$ and $E_v$ ($E_{f,v} \subset E$) indicate the sets of four edges 
that are surrounding the face $f$ and are adjacent to the vertex $v$, respectively.
The code state $|\Psi \rangle$ is defined as the simultaneous eigenstate 
of all stabilizer operators
with the eigenvalue $+1$:
\begin{eqnarray*}
|\Psi \rangle = A_f |\Psi\rangle , \;\; |\Psi \rangle = B_v |\Psi\rangle 
\textrm{ for all} \;\; f,v.
\end{eqnarray*}
As will be seen below, the eigenvalues of these stabilizer operators
are used to diagnose syndromes of errors.

The logical information encoded in the code space
is defined by those operators that are commutable with all elements
in the stabilizer group but are independent from them.
Such operators characterize the degrees of freedom
in the degenerated code space and hence called {\it logical operators}.
The products of $Z$s on any trivial cycle of the lattice
are commutable with stabilizer operators.
But they are also elements of the stabilizer group,
since they are written by the product of all $A_f$s inside the trivial cycle as shown in Fig.~\ref{fig2} (b).
This is also the case for the products of $X$s on any trivial cycle of the dual lattice.
The products of $Z$s ($X$s) on nontrivial cycles wrapping around the torus
on the primal (dual) lattice
give the logical operators,
which we denote by $L_Z$ ($L_X$).
Since the genus of the torus is one,
we can find two pairs of logical operators $(L_X^{(1)}, L_Z^{(1)})$ and 
$(L_X^{(2)}, L_Z^{(2)})$ on the torus as shown in Fig.~\ref{fig2} (c).
Note that the actions of the logical operators on the code space
depend only on the homology class of the logical operators.
Since these logical operators are subject to the Pauli commutation relation $ L_X^{(i)} L_Z^{(j)} = -\delta _{ij}  L_Z^{(j)} L_X^{(i)}$ and $(L_{A}^{(i)})^2=I$,
they represent two logical qubits encoded
in the code space. In general, a surface code defined on a surface of a genus $g$
can encode $2g$ logical qubits.

Let us consider how errors appear on the code state.
We, for simplicity, consider the Pauli operators $X$ (bit-flip error) and $Z$ (phase-flip error) 
act on the code state with an identical and independent distribution.
However, possibility of correcting these two types of errors
guarantees the validity under a general noise model.
While, in the following, we only consider how to correct $Z$ (phase-flip) errors for clarity,
it is straightforward to correct $X$ (bit-flip) errors in the same manner.
Suppose a chain of $Z$ errors occur on the set $C$ of the qubits (see Fig.~\ref{fig2} (d)).
The code state is given by
$ |\Psi ' \rangle =\prod _{i \in E} Z_i ^{ u_i^{C}}  |\Psi \rangle $,
where $u^{C}_i = 1$ if $i \in C$ and $u^{ C}_i = 0$ if $i \notin  C$.
Due to the errors, the state $|\Psi ' \rangle$ is not in the code space anymore.
If $| E_v \cap C|$ is odd,   
\begin{eqnarray*}
B_v |\Psi ' \rangle = -  \left (\prod _{i \in E} Z_i ^{ u_i^{C}}  \right) B_v |\Psi \rangle
=- | \Psi ' \rangle
\end{eqnarray*}
and hence the eigenvalue of the vertex stabilizer $B_v$ is flipped to $-1$.
The error syndrome of a chain $C$ of $Z$ errors is defined 
as a set of eigenvalues $\{ b_v = \pm 1\}$ of the vertex stabilizers $\{B_v\}$.
Similarly the error syndrome of $X$ errors is defined as 
a set of eigenvalues $\{ a_f = \pm 1\}$ of the face stabilizers $\{A_f\}$.
%Suppose a chain $C$ of $Z$ errors occurs on the surface as shown in Fig.~\ref{fig2} (d).
The eigenvalues $b_v=-1$ are obtained at the
boundaries $\partial C$ (grayed squares in Fig.~\ref{fig2} (d)) of the error chain $C$,
since the vertex stabilizers anticommute with the error chain there.
In order to recover from the errors,
we infer the most likely error operator $\bar C_{\rm ML}$
which has the same error syndromes as $C$, i.e., $\partial \bar C_{\rm ML} = \partial C$.
If $C+\bar C_{\rm ML}$ becomes a trivial cycle the error correction succeeds,
since it acts on the code space trivially as a face stabilizer operator.
Otherwise, if the chain $C+\bar C_{\rm ML}$ becomes a nontrivial cycle,
the operator $\prod _{i \in E} Z_i ^{C+\bar C_{\rm ML}}$ is a logical operator.
Thus the recovery operation destroys the original logical information, 
meaning a failure of the error correction.

Suppose errors occur with an independent and identical probability $p$.
(Note that $p$ is a parameter in the posterior probability of the inference problem,
and hence can be different from the actual error probability $p'$.)
The error probability of an error chain $\bar C$ 
conditioned on the $\partial \bar C = \partial C $ becomes
\begin{eqnarray}
P( \bar C | \partial \bar C= \partial C) &=&\mathcal{N} \prod _{i }  \left( \frac{p}{1-p} \right) ^{ u^{\bar C}_i}
\big |  _{ \partial \bar C= \partial C}
=\mathcal{N}'  \exp  \left[-\beta J \sum _{i}  (2u^{\bar C}_i  -1) \right]
\big |  _{ \partial \bar C= \partial C}
\label{posteria}
\end{eqnarray}
where $\mathcal{N}$ and $\mathcal{N}'$ are the normalization factors, $e^{ -\beta J}\equiv \sqrt{p/(1-p)}$.
The most likely error operator $\bar C_{\rm ML}$ conditioned on the syndrome $\partial C$ 
can be obtained by maximizing the posterior probability:
\begin{eqnarray*}
\bar C_{ML}= {\rm arg} \; \max _{\bar C }  \; 
  P( \bar C | \partial \bar C= \partial C)
= {\rm arg} \; \min _{\bar C }  \; 
\sum _{i} u^{\bar C} _{i}  | _{ \partial \bar C= \partial C}.
\end{eqnarray*}
This indicates that the most likely chain $\bar C_{\rm ML}$ is an error chain
that connects pairs of two boundaries with a minimum Manhattan length.
Such a problem can be efficiently solved in classical computer by using the
Edmonds' minimum-weight-perfect-matching (MWPM) algorithm~\cite{Edmon}.
The accuracy threshold with the decoding by the MWPM algorithm
has been estimated to be 10.3\% (MWPM)~\cite{Wang}.

The MWPM algorithm is not optimal for the present purpose,
since we have to take not only the probability of each error $\bar C$ but also the 
combinatorial number of the error chains $\bar{C}$ 
such that $\bar C+C$ belongs to the same homology class
(recall that the action of logical operators on the code state 
depends only on the homology class of the associated chains).
By taking a summation over such a $\bar C$ that belongs to the same homology class,
we obtain a success probability of the error correction
\begin{eqnarray*}
p_{\rm suc} = \mathcal{N}'' \sum _{\bar C + C = \textrm{trivial cycles}}  
\exp  \left[ \beta J \sum _{i}  v_i ^{\bar C} \right]
,
\end{eqnarray*}
where $v _i^{\bar C} = -2 u_i ^{\bar C} +1$ is defined.
In order to simplify the summation over trivial cycles,
we introduce an Ising spin  $\sigma _{i} \in \{ +1, -1 \}$ on each face center (i.e., vertex of the dual lattice) of the lattice (see Fig.~\ref{fig2} (e)).
If $\bar C+C$ a trivial cycle,
there exists a configuration $\{ \sigma _i \}$
such that
$v_{l} ^{\bar C}= v_l ^{C} \sigma _i \sigma _j $ 
with $l$ being the bond between the spins $\sigma _i$ and $\sigma _j$.
The variable $v_{l}$ located on the bond in-between the sites $i$ and $j$ is denoted by $v_{ij}$
hereafter.
By using this fact, the success probability can be 
reformulated as 
\begin{eqnarray*}
p_{\rm suc} &=& \mathcal{N}''' 
\sum _{\{ \sigma _i \}  }  \exp \left[ {  \beta  \sum _{\langle ij \rangle }J_{ij} \sigma _i \sigma _j} \right],
\end{eqnarray*}
where $\mathcal{N}'''$ is a normalization factor, and
$J_{ij} \equiv Jv_{ij}^{C}$.

Now that the relation between quantum error correction and 
a spin glass model becomes apparent;
the success probability of the $Z$ error correction is nothing but the appropriately normalized partition function of the RBIM,
whose Hamiltonian is given by $H= - \sum _{\langle ij \rangle} J_{ij} \sigma _i \sigma _j$.
The location of $Z$ errors represented by $J_{ i j} = -1$ 
corresponds to the anti-ferromagnetic interaction due to disorder,
whose probability distribution is given by
$P( J_{ i  j} ) =(1-p') \delta ( J_{ i  j} -1 )+ p'\delta ( J_{ i  j} +1 )$.
(Recall that the hypothetical error probability $p$ is given independently of 
the actual error probability $p'$.)
The error syndrome $b_v=-1$ corresponds to
the frustration of Ising interactions (see Fig.~\ref{fig2} (e)). 
Equivalently it is also the end point (Ising vortex) of the excited domain wall. 

In order to storage quantum information reliably,
the success probability has to be reduced exponentially 
by increasing the system size.
This is achieved with an error probability $p'$ below
a certain value, so-called {\it accuracy threshold} $p^{\rm th}$.
However, if the error probability is higher than it,
the success probability converges to a constant value.
This drastic change on the function $p_{\rm suc}$
corresponds to a phase transition of the RBIM.
Specifically, in the ferromagnetic phase, quantum error correction succeeds.
(Actually the logical error probability is related to 
the domain-wall free-energy via $- (1/\beta)\ln p_{\rm suc}$.
In the ferromagnetic phase, the domain-wall free-energy
scales like $O(N)$ with $N$ being the vertical or horizontal 
dimension of the system. This supports the exponential 
suppression of logical errors.)
Of course, we should take $p= 1/(e^{2\beta } +1) = p'$, that is,
the actual and hypothetical error probabilities are the same, in order to 
perform a better error correction.
This condition corresponds to the Nishimori line~\cite{Nishimoriline} on the $(\beta ,p)$ phase diagram.
The precise value of the optimal threshold has been calculated 
to be 10.9\%~\cite{Ohzeki}, which is fairly in good agreement with the numerical estimation~\cite{Merz}. 
The threshold value 10.3\% with the MWPM algorithm
corresponds to the critical point at zero temperature,
since the solution of the MWPM algorithm is obtained in the $\beta \rightarrow \infty$ limit.

Recently the surface codes have been also studied on 
general lattices including random lattices~\cite{Ohzeki,Roth,FujiiTokunaga,OhzekiFujii,Al-Shimary}.
Furthermore, the performance analyses in the presence of qubit-loss error
have been also argued~\cite{Loss1,Loss2,OhzekiLoss,FujiiTokunaga},
which corresponds to bond dilution in the RBIM.

The surface code is one example of 
topological stabilizer codes,
whose stabilizer operators are local (i.e., finite-body Pauli products) 
and translation invariant. A complete classification of
topological stabilizer codes is obtained in Ref. \cite{Beni}.
Another well-studied example of topological stabilizer codes
is the color codes \cite{Bombin1,Bombin2},
which are related to the random three-body Ising models.
The performance of the color codes
has been also discussed via the spin glass theory \cite{OhzekiColor,BombinOhzeki}.

An important issue in quantum error correction,
where the knowledge of statistical mechanics seems to be quite useful,
is finding a fast classical decoding algorithm.
Recently several new efficient algorithms have been proposed \cite{Poulin,Fowler,Wootton}.
They are based on a renormalization group algorithm \cite{Poulin},
local search \cite{Fowler}, and parallel tempering \cite{Wootton},
each of which is well studied in statistical mechanics.

In fault-tolerant quantum computation,
we have to take all sources of errors into account,
such as imperfections in syndrome measurements.
Thus, in fault-tolerant quantum computation
the code performances under perfect syndrome measurements
are of limited interest.
For both surface and color codes,
their performances have been investigated under imperfect 
syndrome measurements~\cite{Dennis,Wang,Andrist}.
Interestingly, the inference problems of errors in them
are mapped into three-dimensional random ${\mathbf Z}_2$ gauge theories.
Recently, topologically protected measurement-based quantum
on the symmetry breaking thermal state has been proposed~\cite{FNOM}.
This model is mapped into a correlated random-plaquette ${\mathbf Z}_2$ gauge model,
whose critical point on the Nishimori line is shown to be equivalent to
that of the three-dimensional Ising model. 

More comprehensive studies of 
fault-tolerant quantum computation has been 
taken by considering errors in state initializations, measurements,
and elementary gate operations used in both quantum error correction itself
and universal quantum computation on the code space \cite{Raussendorf06,Raussendorf07a,Raussendorf07b,Landahal}.
Based on these analyses,
physical implementations and quantum architecture designs
have been argued recently \cite{VanMeter,LiBenjamin,FTProb,Cody,FujiiDis,Benjamin1,Benjamin2,Fowler1,Fowler2},
which clarify the experimental requirements for building
fault-tolerant quantum computer.
On the other side, extensive experimental resources have been paid
to achieve these requirements, and 
very encouraging results have already obtained in some experiments, for example, in 
trapped ions \cite{IonQEC,IonHighAccuracy} and superconducting systems \cite{SuperConductingHighAccuracy,SuperConductingQEC}.
Finding a high performance quantum error correction code,
an efficient classical algorithm for decoding, and
a detailed analysis of an architecture under a realistic situation 
will further help an experimental realization of large scale quantum computer.

%%%%%%%%%%%%%%%%%%%%
\section{Statistical models and stabilizer formalism}
In this section, we describe a mapping between
the stabilizer formalism and classical statistical models~\cite{Nest1,Completeness},
which allows us to analyze classical statistical models 
via quantum information and vice versa. 
\begin{figure}[hbt]
\centering
\includegraphics[width=130mm,keepaspectratio,clip]{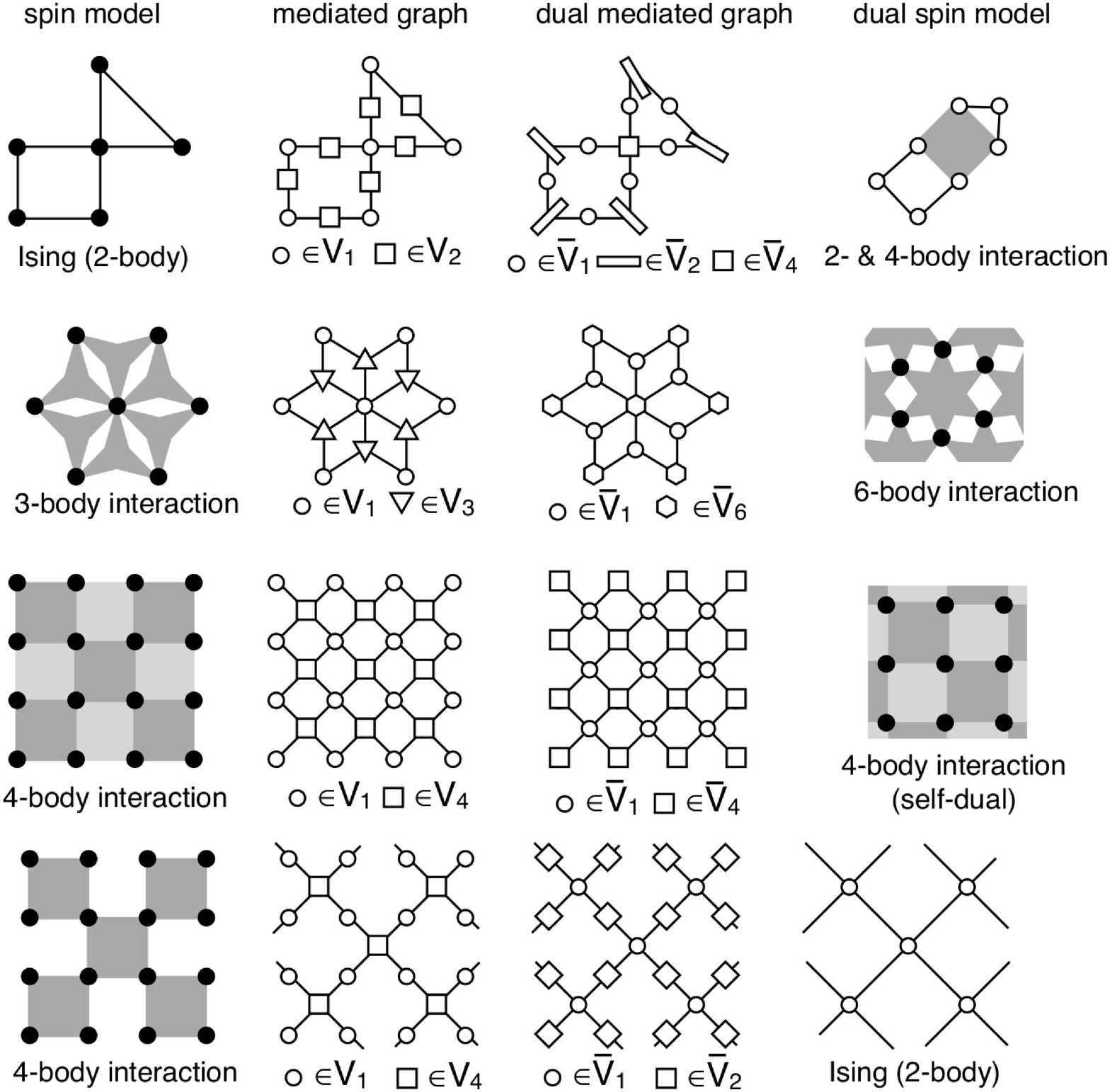}
\caption{From left to right, spin models, corresponding mediated graph states,
dual mediated graph states, and dual spin models.
In the mediated graph states, circles indicate the qubits representing sites 
of spin models. The magnetic fields are given by a product state
defined on these site qubits. The polygons indicate the qubits, which mediate the many-body interactions.
The number of the edges of each polygon is equivalent
to the number of spins engaged in the corresponding interaction.
The coupling strengths of the many-body interactions
are determined by a product state defined on these mediating qubits.
}
\label{fig3}
\end{figure}
More precisely, we will relate
an overlap $\langle \alpha | \Psi_{G} \rangle $ of a product state $| \alpha \rangle \equiv \prod _i | J_i \rangle$, which will be defined later in detail,
and a graph state $|\Psi _{\rm G} \rangle$ to a partition function $\mathcal{Z}(\beta)$
of a spin model associated with interactions $J_i$:
\begin{eqnarray*}
\mathcal{Z}(\beta) = \frac{1}{2^{|V_1 | /2}} \langle \alpha |  \Psi _{G}\rangle ,
\end{eqnarray*}
where $|V_1|$ denotes the number of sites in the spin model.
Since the right hand side can be efficiently evaluated by
using quantum computer, the overlap mapping gives us 
an algorithm to calculate the partition functions.
Below we explain a derivation of the overlap mapping in detail.

We consider a classical statistical model on a lattice $L$:
\begin{eqnarray*}
H =  \sum _i J_i s_i  + \sum _{\langle i,j \rangle}J_{i,j}  (s_i \oplus s_j) + \sum _{\langle  i,j,k  \rangle} J_{i,j,k} (s_i \oplus s_j \oplus s_k )+ \dots + \sum _{\langle i_1, i_2, ..., i_m \rangle} J_{i_1,i_2,...,i_m} \left( \bigoplus _{j=1}^{m} s_{i_j} \right), 
\end{eqnarray*}
where $s_i \in \{ 0,1\}$ is the $i$th Ising spin variable,
$ J_{ i_1,i_2,...,i_m}$ is the coupling constant for the $m$-body interaction on $i_1$th, ..., and $i_m$th spins,
and $\bigoplus$ indicates the addition modulo 2.
We can choose any lattices and interaction geometries as shown in Fig.~\ref{fig3}.
Let us denote the number of $j$-body interactions by $n_j$.
Specifically, $n_1$ is the number of sites.
Next we relate the classical statistical model
to another graph. 
This graph consists of $m$ sets of vertices, $V_j$ ($j=1,...,m$),
each of which has $n_j$ vertices. The total number of vertices 
on the graph is denoted by $n\equiv \sum _{j=1}^{m} n_m$. 
The set $V_1$ and $V_j$ ($j \neq 1$) 
represent information about the sites and the $j$-body interactions, respectively. 
The graph associated with the classical model on a lattice $L$,
which we call the mediated graph,
is defined such that if there is a $j$-body interaction among the sites $i_1, i_2, ...$, and $i_j$,
the corresponding vertices in $V_1$ and $V_j$ 
are connected as shown in Fig.~\ref{fig3}.
Then we consider the graph state on the mediated graph.
By using the definition given in Sec. 5,
the mediated graph state $| \psi _G \rangle$can be written by
\begin{eqnarray*}
|\psi _{G} \rangle =       \prod _{(i,j)\in E}  \Lambda (Z)_{i,j}  |+ \rangle ^{ \otimes n}
\Leftrightarrow
\left( \prod _{i \in  \bigcup _{j=2}^{m} V_j} H_i \right) |\psi _G \rangle
&=&  \prod _{(i,j)\in E \text{ where } i \in n_1}  \Lambda (X)_{i,j}  |+ \rangle ^{ \otimes |V_1| } |0 \rangle ^{ \otimes n- n_1}
\\
&=& \frac{1}{2^{ n_1/2}}\sum _{\{s_i \} } \bigotimes _{i} | s_i \rangle    \bigotimes  _{\langle i,j \rangle } |s_i \oplus s_j \rangle \otimes
.... \bigotimes _{\langle i_1, ... , i_m \rangle}  | \bigoplus _{j=1}^{m} s_{i_ j} \rangle
\end{eqnarray*}
where we used that facts that $H_j \Lambda (Z)_{i,j} H_{j}= \Lambda (X)_{i,j} $
and $|+\rangle = H|0\rangle$.
The summation $\sum _{\{ s_i \}}$ is taken over all configurations of $\{s_i \}$.
The mediated graph state does not have any information about
the coupling constants. Thus we define a product state
$\displaystyle \bigotimes _{i}  | J_i \rangle \bigotimes _{\langle i,j \rangle} | J_{ij} \rangle \otimes ... \otimes 
\bigotimes _{\langle i_1,...,i_m \rangle} |J_{ i_1,..., i_m } \rangle$,
where $| J \rangle \equiv |0 \rangle +e^{ -\beta J } |1\rangle$.
Then we take the inner product between the mediated graph state with Hadamard gates
and the product state:
\begin{eqnarray}
\displaystyle \bigotimes _{i \in V_1}  \langle  J_i | \bigotimes _{\langle i,j \rangle \in V_2} \langle J_{ij}| \otimes ... \otimes 
\bigotimes _{\langle i_1,...,i_m \rangle \in V_m} \langle J_{ i_1,..., m } | \left( \prod _{i \in \bigcup _{j=2}^{m} V_j } H_i \right) |\psi _G \rangle = \frac{1}{2^{ |V_1|/2}}\sum _{s_i} e^{ - \beta H} =\frac{1}{2^{ |V_1|/2}}Z(\beta).
\label{VDB}
\end{eqnarray}
This is the VDB (Van den Nest-D{\"u}r-Briegel) overlap mapping obtained in Refs.~\cite{Nest1,Completeness}
(an extension to $q$-state spin models is straightforward~\cite{Completeness}).
The VDB overlap mapping has a great potential to understand 
both quantum information and statistical mechanics.
First of all, the overlap between the product and graph states
can be regarded as an MBQC
as given in Eq. (\ref{MBQC}). Thus VDB overlap mapping helps us to find a new quantum algorithm
that calculates partition functions~\cite{Nest2,algorithm1,algorithm2}. If a spin model
with an appropriately chosen lattice geometry  and coupling constants
allows universal MBQC,
evaluation of the corresponding partition function
is  BQP-complete, that is,
as hard as any problem that quantum computer can solve.
Furthermore, by using universality of MBQC,
we can relate different statistical models~\cite{Completeness,gauge1,gauge2,U1gauge}.
For example, the partition functions of all classical spin models are equivalent 
to those of the 2D Ising models with appropriately chosen complex coupling constants and magnetic fields on polynomially enlarged lattices~\cite{Completeness,Karimipour1}.
On the other hand, if a quantum computation is related to 
an exactly solvable model, the corresponding quantum computation is classically simulatable~\cite{Nest2}.
MBQC
on certain types of graphs, such as tree graphs, has been known to be classically simulatable~\cite{Tree,Universal1,Universal2}.
This also provides a good classical algorithm to evaluate the partition function~\cite{Nest2}. 

Very recently, another overlap mapping has been developed in Ref.~\cite{NestDur},
which has the following forms:
\begin{eqnarray*}
\mathcal{Z}_{3{\rm -Ising}} =\frac{1}{2^{|V_1|}} \langle  \Psi _{\rm color} | A | \Psi _{\rm color}\rangle.
\label{new_map}
\end{eqnarray*}
The left hand side is a partition function of an
Ising model with only three-body interactions.
In the right hand side, 
$|\Psi _{\rm color}\rangle$ is the color code states \cite{Bombin1,Bombin2},
and 
$A\equiv \bigotimes _i (Z S^{\dag} H D_i H S)$ with $D_i$ being 
a $2\times 2$ diagonal matrix.
In comparison with the VDB overlap mapping,
the mapping Eq. (\ref{new_map}) improves the scale factor exponentially.
Since the overlap $\langle  \Psi _{\rm color} | A | \Psi _{\rm color}\rangle$
is the mean value of an observable $A$ with respect to $|\Psi _{\rm color}\rangle$,
its evaluation can be easily obtained in quantum computer.
Furthermore, the overlap can be rewritten as
\begin{eqnarray*}
\langle  \Psi _{\rm color} | A | \Psi _{\rm color}\rangle 
= \langle \Psi | \bigotimes _i D_i  | \Psi \rangle,
\end{eqnarray*}
where $|\Psi \rangle$ is a stabilizer state.
Recently, it has been found that
the right hand side can be simulated efficiently in classical computer with a polynomial overhead,
which uses classically simulatability of a restricted type of 
quantum computation, i.e., Clifford circuit \cite{Nest11}. 
This provides an efficient classical algorithm, taking a detour,
to evaluate partition functions of three-body Ising models.

As a final topic, we demonstrate a duality relation between
two classical spin models by using the VDB overlap mapping.
We define a set of vertices $\bar V_{j} := \{ i | i\in V_1 \textrm{ and connected to $j$ vertices on the mediated graph} \}$ and $\bar V_1 = V_2 \cup .... \cup V_m$.
Then, Eq. (\ref{VDB}) is reformulated as
\begin{eqnarray}
&&
\bigotimes _{\langle i,j \rangle \in \bar V_2} f_{i,j} \langle \tilde J_{i,j} | ...
 \bigotimes _{ \langle i_1,..., i_j \rangle  \in \bar V_j}  f_{{i_1, ..., i_j }} \langle \tilde  J_{i_1, ..., i_j } | 
\bigotimes _{i \in \bar V_1} f_i \langle \tilde J_{i} | 
\left( \prod _{i \in  \bigcup _{j=2}^{m} \bar V_j } H_i \right) |\psi _G \rangle 
\nonumber \\
&&=\frac{1}{2^{|\bar V_1|/2}} \prod _{i \in \bar V_1 } f_1 \prod _{ \langle i,j  \rangle \in \bar V_2} f_{i,j} ... \prod _{\langle i_1, ..., i_m \rangle \in \bar V_m} f_{i_1, ..., i_m} \tilde Z(\beta ) 
\label{VDB2}
\end{eqnarray}
Here we defined $f_{i_1, ..., i_j } \langle \tilde  J_{i_1, ..., i_j } | = \langle  J_{k(i_1, ..., i_j )} | H$ 
with site $k(i_1, ..., i_j ) \in V_1$ being connected to sites $i_1, .. , i_j$,
and $f_{\bar k (i_1, ..., i_j)} \langle \tilde J_{\bar k (i_1, ..., i_j)} | = \langle J_{i_1, ..., i_j} | H$
with site $\bar k(i_1, ..., i_j ) \in \bar V_1$ being connected to sites $i_1, .. , i_j$,
where $f_{i_1, ..., i_j }$ and $f_{\bar k (i_1, ..., i_j)}$.
These $\tilde  J_{i_1, ..., i_j }$, $f_{i_1, ..., i_j }$, $\tilde J_{\bar k (i_1, ..., i_j)} $, and $f_{\bar k (i_1, ..., i_j)}$ are specifically given by
\begin{eqnarray*}
e^{-\beta \tilde  J_{i_1, ..., i_j }} = \tanh ( \beta J_{k(i_1, ..., i_j )}/2), \;\;
%		  (1 - e^{ - \beta J_{k(i_1, ..., i_j )}})/(1 + e^{ - \beta J_{k(i_1, ..., i_j )}}),
f_{i_1, ..., i_j } = (1 + e^{ - \beta J_{k(i_1, ..., i_j )}})/\sqrt{2},
\\
e^{- \beta \tilde J_{\bar k (i_1, ..., i_j)}} = \tanh ( \beta J_{i_1, ..., i_j }/2), \;\;
f_{\bar k (i_1, ..., i_j )} = (1 + e^{ - \beta J_{i_1, ..., i_j }})/\sqrt{2}.
\end{eqnarray*}
Up to the constant factor, 
R.H.S. of Eq. (\ref{VDB2}) reads the partition function of the dual model,
which corresponds to the mediated graph with the set of vertices $\bar V_j$ (see Fig.~\ref{fig3} for examples).
For example, the two-body Ising model with magnetic fields on the square lattice
is the dual of the four-body Ising model of a checker board type 
with magnetic fields as shown in Fig.~\ref{fig3} (the forth row).
A four-body Ising model on all plaquettes with magnetic fields 
is self-dual as shown in Fig.~\ref{fig3} (the third row).  
While we only considered the duality relations between spin models on
two dimensions here, it is straightforward to extend them to general cases.

\section{Conclusion}
In this review, we give a pedagogical introduction
for the stabilizer formalism.
Then we review two interdisciplinary topics,
the relations between the surface code and the RBIM,
and the mapping between the stabilizer formalism and the partition function of classical spin models.
Both of two are quite important topics in quantum information science;
one provides a basis for building fault-tolerant quantum computer,
the other provides a new quantum algorithm to evaluate 
the partition functions of statistical mechanical models.

These topics mentioned here are a small fraction of 
interdisciplinary fields between quantum information science 
and statistical mechanics, or more generally, condensed matter physics. 
The coding theory based on the stabilizer quantum error correction codes
provides an important clew to understand topological order \cite{Kitaev,BeniTopo}.
There is a quantum algorithm that calculates the Jones and Tutte polynomials~\cite{AharonovJones,AharonovJonesHard,AharonovTutte},
which are equivalent to the partition functions of Potts models
with complex coupling constants.
It has been shown that the class of additive approximations of Jones or Tutte polynomials
at certain complex points within specific algorithmic scales is BQP-complete.  
There is a classically simulatable class of quantum computation,
the so-called match gates \cite{Match_gate1,Match_gate2,Match_gate3,Match_gate4},
which is based on the exact solvability of free-fermion systems.
Here we did not addressed quantum annealing \cite{NishimoriKadowaki} 
or adiabatic quantum computation \cite{Farhi},
where a ground state configuration of a statistical mechanical model
is obtained by adiabatically changing parameters of another 
quantum Hamiltonian.
The quantum annealing or adiabatic quantum computation 
are thought to be easy for a physical implementation \cite{D-wave},
while there are still ongoing debates \cite{IBM} on whether or not the current experimental quantum annealing \cite{D-wave} utilizes 
genuine quantumness. 
The adiabatic model has been shown to be 
equivalent to the standard model of quantum computation in an ideal situation~\cite{Aharonov1,Aharonov2}. However, the equivalence in an ideal situation
does not mean that a fault-tolerant theory in one model
automatically ensures fault-tolerance in the other model.
Thus fault-tolerance of the adiabatic model 
against all sources of experimental imperfections 
has to be addressed \cite{Child,Lidar}.

The correspondences between quantum information science
and other fields including statistical mechanics 
are continuing to be discovered, and
the complete understanding of both fields in the same language 
is being obtained.
The interdisciplinary interactions will 
further deepen understanding of both fields of physics,
and will lead us to an emerging new field.

\section*{Acknowledgments}
Grant-in-Aid for Scientific Research on Innovative Areas No. 20104003.

\end{document}